\documentclass[11pt]{article}  

\usepackage{amsmath}
\usepackage{amssymb}
\usepackage{amsthm}
\usepackage[USenglish]{babel}
\usepackage[utf8]{inputenc}
\usepackage[T1]{fontenc}
\usepackage{color}
\usepackage{graphicx}
\usepackage{enumerate}
\usepackage{url}
\usepackage{authblk}
\usepackage{xspace}
\usepackage{stmaryrd}
\usepackage{booktabs}
\usepackage{subcaption}

\usepackage{tikz}
\usetikzlibrary{decorations.pathreplacing}

\usepackage{algorithm}
\usepackage[noend]{algpseudocode}
\newcommand{\TO}{\textbf{to}\xspace}

\newtheorem{definition}{Definition}

\newcommand{\minlogapa}{ {\sc MinLogGapA}\xspace}
\newcommand{\recbis}{{\sc RecBis}\xspace}
\newcommand{\dual}{{\sc Dual}\xspace}
\newcommand{\bimlogapa}{{\sc BiMLogGapA}\xspace}
\newcommand{\extshingle}{{\sc SimRef}\xspace}

\begin{document}

\title{Compressing bipartite graphs with a dual reordering scheme}

%\shorttitle{Compressing bipartite graphs with a dual reordering scheme}
%\shortauthorlist{M. Danisch, I. Panagiotas, L. Tabourier}

\author{Maximilien Danisch}
\author{Ioannis Panagiotas}
\author{Lionel Tabourier\thanks{corresponding author: \texttt{lionel.tabourier@ens-lyon.org}}}

\affil{Sorbonne Université, CNRS, LIP6, F-75005 Paris, France}

\date{}

\maketitle

\begin{abstract}
{
In order to manage massive graphs in practice, it is often necessary to resort to graph compression, which aims at reducing the memory used when storing and processing the graph.
Efficient compression methods have been proposed in the literature, especially for web graphs. 
In most cases, they are combined with a vertex reordering pre-processing step which significantly improves the compression rate.
However, these techniques must be adjusted to the type of graphs under consideration.
In this paper, we focus on the class of bipartite graphs and adapt the vertex reordering phase to their specific structure by proposing a dual reordering scheme.
By reordering each group of vertices in the purpose of minimizing a specific score, we show that we can reach better compression rates. 
We also suggest that this approach can be further refined to make the node orderings more adapted to the compression phase that follows the ordering phase.
}
% keywords
{
Bipartite graphs; Graph algorithms; Graph compression; Vertex Ordering.
}
\\
%2000 Math Subject Classification:
\end{abstract}

%%%

\section{Introduction \label{sec:intro}}

Many real-world systems are adequately represented by graphs, as they allow to model interacting entities.
Among many examples, graph representations are used to describe the world wide web as webpages connected by hyperlinks, social networks where accounts are connected by a friendship or a follower relationship, human activity on online platforms, words occurring in a same sentence, or even biochemical reactions between proteins.
As data is generated faster and in larger quantities than ever before, we are led to handle massively increasing graph sizes. 
For example, graphs from social networks such as Facebook or Twitter demand up to terabytes of RAM to be loaded in main memory.
It is impossible to process them with standard computers, making the retrieval and analysis of relevant information more challenging.
In this context, designing methods to produce compact representations of the information contained in a graph has emerged as an important research question, named {\em graph compression}.
The compression can be either  {\em lossy} or {\em lossless}, depending on whether part of the information is left out of the representation or not.
Here, we are interested in lossless compression.

Very efficient methods for lossless graph compression have been proposed in the literature~\cite{bv04,vccg20}.
These methods have been designed primarily to be applied to web graphs, \textit{i.e.} graphs where nodes are URLs and directed edges represent hyperlinks. In this case, nodes are naturally ordered according to the lexicographic order of the URLs.
It was observed that this natural vertex ordering favors good compression rates notably because it satisfies two properties that compression techniques exploit: {\em locality} and {\em similarity}~\cite{bv04}.
The first term means that a vertex is mainly connected to vertices with a close index, while the second means that vertices which share many neighbors have close indexes.
Unfortunately, other types of networks do not necessarily exhibit such an adequate natural vertex ordering, leading to much less efficient compressed representations.
That is why vertex reordering techniques have emerged as an essential preprocessing phase to achieve high graph compression rates~\cite{chklmpr:09,borsv:11,dkkops:16}. 
These techniques seek to find an ordering that satisfies as much as possible the properties of locality and similarity, as well as any other property which could benefit the subsequent compression phase.

In this work, we focus on the problem of lossless compression of bipartite graphs and especially on the reordering phase of the process.
Bipartite graphs are appropriate models to represent systems where relations connect two different kinds of entities: users consuming contents on an online platform, individuals and groups to which they are affiliated, indexes referencing pages, etc.
We utilize the fact that both groups of nodes play very different roles in the graph and propose  reordering the two groups independently using two distinct objectives, rather than optimizing a single objective as is usual.
These objectives are defined to be consistent with the logic of the compression methods that is applied.

%Experimentally, we will demonstrate that our proposed dual reordering scheme is more efficient for bipartite graphs than the standard single reordering approach. 

The rest of the paper is organized as follows.
In Section~\ref{sec:back} we provide the background and summarize related work.
%
%In Section~\ref{sec:refing} we discuss modification to the standard referencing schemes of existing solutions~\cite{bv:04,vccg:20}.
%
Then in Section~\ref{sec:ordings}, we present the dual reordering scheme that we propose to compress bipartite graphs efficiently.
Section~\ref{sec:exps} is dedicated to the description of the experimental protocol and results that we have obtained, which show that the scheme is indeed more efficient than state-of-the-art reordering methods for this purpose.
We present in Section~\ref{sec:discus} some leads to develop this reordering method further.% before briefly concluding in Section~\ref{sec:conc}.

\section{Background and related work \label{sec:back}}

\paragraph{Preliminaries.}
We first define the basic vocabulary needed to describe the problem of graph compression.
Let $G=(V,E)$ be an undirected graph where $V$ is the set of vertices and $E$ the set of edges.
We call $u,v \in V$ neighbors in $G$ if the edge $(u,v)$ exists in $E$.
The {\em neighborhood} of a vertex $u$, denoted $ \mathcal{N}(u)$, is the set of its neighbors and the degree of $u$, $d_u$, is the size of this neighborhood.
%
%A {\em path} in $G$ is a set of edges $(u_0,v_0),\ldots, (u_k,v_k)$ from $E$ such that $v_i=u_{i+1}$ for every $i < k$.
%The parameter $k$ denotes the {\em length} of the path and a path is {\em odd} or {\em even} based on the parity of $k$.
%A path forms a {\em cycle} if the starting and ending vertices are the same i.e., $u_0$ is equal to $v_{k}$.
A graph is {\em bipartite} if $V$ can be partitioned into two disjoint sets $V_\top$ and $V_\bot$ such that $V=V_\top \cup V_\bot$ and every edge in $E$ contains one vertex from $V_\top$ and one from $V_\bot$. 
$V_\top$ and $V_\bot$ are called respectively \textit{top} and \textit{bottom} nodes.
%
%A directed graph $G$ is a graph such that every edge has directions, i.e., we denote edges by $u \rightarrow v$. 
%In this case, we call $v$ an {\em out-neighbor} of $u$ and $u$ an {\em in-neighbor} of $v$.
%The {\em out-neighborhood} or {\em in-neighborhood} of a vertex is the set of its out-neighbors or in-neighbors respectively.
%A directed acyclic graph (DAG) is a directed graph where the edges are oriented in such a way that the graph does not contain any cycles.
%
As previously mentioned, graph compression is tightly linked to the node indexation of the graph.
%
%Indeed, vertex reordering has been shown to be an important pre-processing phase for efficient graph compression~\cite{chklmpr:09,borsv:11,dkkops:16}.
%
We define an ordering of $V$ as an injective function $\pi : V \rightarrow \{ 1, \ldots , \vert V \vert \}$, \textit{i.e.}, a renumbering of the vertices of $V$.
%Two vertices $u,v$ are called consecutive in $\pi$ if $\vert \pi(u) - \pi(v) \vert =1$. 
Vertex $u$ precedes $v$ if $\pi(u) < \pi(v)$ and follows $v$ if $\pi(u)> \pi(v)$.

Note that graphs built from real-world data are known to be sparse. 
A straightforward way to store efficiently sparse graphs in memory is to represent them as adjacency lists \textit{i.e.}, the lists of neighbors of each node in the graph.
This format is the starting point of efficient lossless graph compression methods, which then use a combination of techniques, that we discuss below, to improve upon them.

\paragraph{Compression methods for web graphs.}
In the case of web graphs, 
Boldi and Vigna's \textit{WebGraph} framework~\cite{bv04} identified several lossless compression techniques that have allowed to reach high compression rates, which are still competitive with current state-of-the-art methods.
Indeed, the compressed representation of such graphs requires as few as 2--3 bits per edge
%
%\footnote{}
-- note that lossless compression is usually empirically measured in bits per edge, as the total size of the graph is not necessarily easy to interpret.
The general idea underlying this framework is to take advantage of two central characteristics of web graphs.
The first one, named {\em locality}, supposes that connected vertices have relatively close indexes.
In lexicographically ordered web graphs, it is usually the case as the source and target of a hyperlink are often part of a same domain name. 
The second one, {\em similarity}, means that similar vertices (nodes which share a large subset of their neighbors) appear close to each other in the ordering. 
This is also the case in web graphs because two pages of a same domain tend to have close index according to the lexicographic order and these two pages commonly have very similar navigational links.

Among the compression techniques developed for WebGraph, Boldi and Vigna defined the {\em zeta encoding}, which is a universal code\footnote{A universal code is a prefix code such that the expected lengths of the words are within a constant factor of the expected lengths of the words using the optimal code.} especially designed for graphs which encodes a sequence of consecutive integers as intervals (\textit{i.e.}, represented only with two values).
%
%nd techniques which take advantage of two central characteristics of web graphs
%
They have also incorporated a widely-used compression technique called {\em delta encoding}~\cite{chklmpr:09,vccg20}: when storing a list of sorted integers $u_1, \ldots, u_k$, delta encoding consists in representing it as a list of consecutive gaps \textit{i.e.}, $u_1, (u_2 - u_1 -1), \ldots, (u_{k}-u_{k-1}-1)$.
Because these gaps are likely to be much smaller than the integers of the original list, storing them with variable-length quantity encoding schemes usually requires less space.
This is particularly efficient if the neighbors of a node have close index, which is consistent with the locality principle.
Another crucial technique for graph compression -- especially in the perspective of our work -- is {\em referencing}.
Referencing allows a vertex $u$ to select one of its predecessors $v$ within a fixed window size as its reference, and store their common neighbors implicitly.
Practically, one keeps for the node $u$ a $0/1$ bit array $b$ of size $d_v$ where $b[i]$ is $1$ if $u$ and $v$ share the $i$th neighbor of $v$ and $0$ otherwise.
The neighbors of $u$ which are not neighbors of its reference $v$ (called residual neighbors) are stored explicitly after the bit array $b$.
When $u$ and $v$ have many common neighbors, such a representation is more efficient than explicitly storing all of $u$'s neighbors.

A few other approaches can be mentioned regarding web graph compression.
On the one hand, some of them develop on Boldi and Vigna's framework.
For instance, Grabowski and Bieniecki~\cite{gabi:14} handle referencing in an alternative way: a vertex $u$ can only reference its immediate predecessor $v$, but each copied neighbor of $v$ is allocated more than one bit, so as to portray more information about the neighborhood of $u$.
%
%for instance~\cite{grabowski2010tight}, where the authors propose to improve the referencing scheme by adding the possibility for a node to refer to several nodes.
%
Also, Liakos \textit{et al.}~\cite{lipasi:14} store the denser diagonal part of the adjacency matrix of the graph separately  for efficiency purposes and resort to using the WebGraph framework for the rest of the graph.
Recently, Versari \textit{et al.}~\cite{vccg20} proposed Zuckerli, a new software for compression that incorporates several improvements to the WebGraph framework.
These improvements notably consist in employing a hybrid encoding scheme for storing integers and improving the referencing algorithm.
They have shown empirical improvements ranging from $10\%$ to $20\%$ on many instances over WebGraph and other methods.
On the other hand, some works are based on schemes which are radically different from %Boldi and Vigna's, 
WebGraph, such as Brisaboa \textit{et al.}~\cite{bilana:09} which introduces $k^2$-trees to compress web graphs.
Their idea is to recursively partition the adjacency matrix of a graph into $k^2$ parts and stop when a submatrix consists solely of 0 or 1.
By storing the resulting tree in an adequate structure~\cite{derara:06}, they managed to achieve efficient compression rates.
%
%it has been shown to achieve efficient compression rates and has brought about a variety of extensions.
%
Another approach is using {\em virtual nodes} to represent frequently appearing structures in the graph.
For instance, Buehrer and Chellapilla~\cite{buce:08} replace bicliques with virtual nodes to encode them more efficiently, then the remaining edges are encoded using standard web graph encoding scheme, such as the ones mentioned earlier, whereas Rossi and Zhou~\cite{rozh:18} represent cliques with virtual nodes.
Claude and Navarro~\cite{cana:10} take inspiration from grammar-based compression to adapt the Re-Pair~\cite{lamo:00} algorithm to graphs: Re-Pair continuously replaces a pair of symbols (here, vertices) that appear together frequently in adjacency lists with a new symbol.
Aside from the method discussed above, Grabowski and Bieniecki~\cite{gabi:14} proposed another compression scheme, which separates vertices into blocks and ``merges'' the neighborhoods of the vertices in each block so as to remove duplicate information.
%
%\ipadd{
%    -To add : S. Grabowski and W. Bieniecki. quick description 
%    -Re-pair and mention generally about frequency %%% lt: I've dropped Re-pair from the paper... 
%Brisaboa et al.~\cite{bilana:09} introduce $k$\^{}2-trees to compress web graphs.
%They recursively partition the adjacency matrix of a graph into $k$\^{}2 parts and stop when a submatrix consists exclusively of ones or zeros. The information is stored as a tree adaptly~\cite{derara:06} encoded for efficiency.
%
%Buehrer and Chellapilla~\cite{buce:08} use the fact that bi-cliques (bipartite subgraphs where one set points fully to the other set) appear frequently in web-graphs.  
%They replace bi-cliques in the graph with virtual nodes that can be encoded more efficiently and encode the remaining edges using aforementioned techniques such as delta-encoding or integer encoding schemes like $\zeta$-codes and achieve better rates than WebGraph.}
%}

A more extensive coverage of various approaches to lossless compression on web graphs can be found in a survey by Besta and Hoefler~\cite{beho:18}.

\paragraph{Inverted index compression.} 
Among other families of data that have attracted much interest for lossless compression purposes, inverted indexes are essential data structures for the efficient implementation of many information retrieval tasks.
They are used to index large collections of documents in the form of an ordered list of IDs, corresponding to the documents where a specific word appear.
% thus fulfilling the opposite role as the one that an index does.
%
While this description stresses the asymmetry between the words (that we call \textit{queries} for generality) and the documents (that we call \textit{data}), an inverted index can be described and stored in the same way as a bipartite graph, where a node either represents a piece of data or a query.
Note that inverted indexes may be enriched with additional information, but we are not interested in this aspect here as we focus on what the bipartite structure  brings.
%but here we are interested in bipartite graphs and thus we do not focus on this aspect.

According to a recent review~\cite{pibiri2020techniques}, the inverted index compression process can be split into three main parts: i) compressing a single integer, ii)  compressing a list of integers and iii) compressing many lists together.
These steps can be mapped to the techniques described in Boldi and Vigna WebGraph framework.
i) The combination of delta and zeta encodings aim at compressing single integers. This combination is particularly efficient because zeta-encoding is well-suited for small integers, while delta encoding ensures to have small integers if the graph has high locality.
Yet other integer encoding methods are available such as Elias gamma and delta codes~\cite{Elias:75} or Rice code~\cite{rice1971adaptive} as well as their many offsprings. 
ii) Compressing a list of integers usually relies on clustering consecutive integers to use a summarized representation of the cluster. 
For that purpose, WebGraph uses notably a variation of binary packing, because their coding method implies the existence of long sequences of 0 and 1.
Here again, other methods are possible, such as entropy coding techniques, among which Huffman coding~\cite{moffat2019huffman} or Arithmetic coding~\cite{moffat1998arithmetic} are probably the most famous.
iii) The referencing approach used in WebGraph is one of the usual ways to compress an ensemble of similar enough lists. Among other possibilities, we point to Pibiri \textit{et al.}~\cite{pibiri2019fast} who proposed a dictionary based approach that yields good results for this part of the process.

In short, we observe that the principle of the methods developed for web graphs are similar to those used for inverted index compression.
Inverted indexes can be seen as bipartite graphs, which makes these techniques relevant for our problem.
Moreover, in the case that we consider, we will have the liberty to use nodes indifferently as queries or data.

\paragraph{Vertex reordering.}
Other types of networks have specific structural properties which can be used to achieve efficient compression.
For instance, in the case of social networks, Chierichetti \textit{et al.}~\cite{chklmpr:09} exploit the fact that these networks often exhibit high {\em reciprocity}~\cite{b:15}, meaning that if there is a directed edge from node $i$ to node $j$, there is a high probability that there is also one from $j$ to $i$.
So it is possible to improve the compression by simply signaling that a link is reciprocal and discarding the explicit reciprocal link.
%
%Note that this trick has a drawback as it slows down the decompression process.
%
Importantly, the authors of~\cite{chklmpr:09} also notice that the methods described for web graphs usually works well because the data collection ordering respects to some extent similarity and locality.
%
%This is for instance the case of the ordering obtained with a lexicographic crawling for web graphs.
%
Unfortunately, the collection ordering with other data -- and particularly social data -- depends on the crawling procedure and it is not guaranteed to be as efficient in terms of locality and similarity.
Consequently, they suggest that before applying the usual compression techniques, one should reorder the vertices of the graph to have as much as possible similarity and locality.

An efficient delta encoding supposes that sibling nodes -- i.e. nodes which are neighbors of a same node -- have close indexes, reciprocally, good referencing demands that nodes with close indexes share many neighbors.
The similarity principle can ensure these two properties.
%
%Delta encoding as well as referencing point to the fact that compression benefits from similarity and locality.
%
These observations highlight the fact that vertex ordering plays an important role in graph compression.
Besides, as the vertex reordering phase is done only once in general, the question of additional reordering time cost is usually irrelevant since it can be done offline.
This contrasts with the compression and decompression times which are independent of the vertex ordering method. 
In this spirit, Apostolico and Drovandi~\cite{aodr:09} proposed GBFS which reorders the vertices using breadth-first search in a preliminary step to improve compression.
Also, Boldi \textit{\textit{et al.}}~\cite{borsv:11} suggested an approach called LLP inspired by community detection algorithms, which partitions vertices into communities and puts vertices of a same community in consecutive positions in $\pi$.
Several other reordering methods relying on a similar community-based strategy are discussed in~\cite{beho:18}.

These techniques aim at optimizing the quality of the vertex ordering in the perspective of future processings, the first of which being delta encoding.
It is therefore natural to express this issue as an optimization problem on orderings.
Chierichetti \textit{\textit{et al.}}~\cite{chklmpr:09} introduced the \minlogapa problem: given a graph $G$, \minlogapa seeks the ordering $\pi$ that minimizes the objective function:
\begin{equation}\label{eq:minloga}
  \sum_{v \in V}  f_{\pi}(v,\mathcal{N}(v)) = \sum_{v \in V} \sum_{i=1}^{d_v -1} \log{( \pi(u_{i+1}) -\pi(u_{i}))}.
\end{equation}
where vertices $u_1$ to $u_{d_{v}}$ are the neighbors of vertex $v$ sorted by increasing value of $\pi$ such that any consecutive difference is positive.
By minimizing the \minlogapa objective function, one minimizes the gaps between neighbors, and it is expected that the delta encoding described above will be more efficient.
%
%It is related to earlier optimization problems, especially {\sc MinLogA} which seeks an ordering to minimize $\sum_{(u,v) \in E}  \log{\vert \pi(u) - \pi(v) \vert}$. % potentiellement enlevable

In~\cite{chklmpr:09}, the authors propose a heuristic called {\em shingle}, which computes a fingerprint of the neighborhood per node and re-orders nodes with similar fingerprints near each other.
We describe it with some more details as we use its principles later in this work.
Supposing that $\sigma$ is a random permutation of a set, then the smallest element in the set according to $\sigma$ is defined as its {\em shingle}.
It has been shown~\cite{broder2000min}  that the Jaccard coefficient $ J(A,B) = \vert A \cap B \vert / \vert A \cup B \vert$ is the probability that the sets $A$ and $B$ have the same shingle.
So a shingle -- or several shingles generated with different permutations, depending on the desired precision -- can be used to fingerprint the set of neighbors of the nodes in the purpose of evaluating approximate Jaccard index.
As the Jaccard index is a measurement of the similarity of two sets, it is used to evaluate the overlap between the neighborhoods of pairs of nodes.
In other words, a shingle is used to approximate $ J(\mathcal{N}(u),\mathcal{N}(v)) = \vert \mathcal{N}(u) \cap \mathcal{N}(v) \vert / \vert \mathcal{N}(u) \cup \mathcal{N}(v) \vert$, which in turn measures the overlap between the neighborhoods of nodes $u$ and $v$.
The shingle re-ordering technique consists in ordering nodes in such a way that consecutive nodes have a high Jaccard coefficient, thus leading to relatively high level of similarity between nodes.
Note that in practice, Chierichetti \textit{et~al.}~\cite{chklmpr:09} use a combination of hash functions instead of random permutations to approximate Jaccard coefficients.
It is a fast reordering method and yields an efficient minimization of the objective function of the \minlogapa problem in comparison to the natural ordering (\textit{i.e.}, the vertices order as produced by the data collection process).
They suggested, but did not prove, that the \minlogapa problem is NP-hard.
It was later shown by Dhulipala \textit{et al.}~\cite{dkkops:16} and they also proposed an efficient recursive bisection heuristic to tackle it practically.
We describe this heuristic in more details below as we make use of it in this paper.

\paragraph{The case  of recursive bisection reordering.}
\label{sssec:recbis}
Dhulipala \textit{et al.}~\cite{dkkops:16} method targets compression of undirected graphs in general, however it is based on optimizing an objective for inverted indexes.
Indeed, they first transform the original graph $G_{u}=(V,E)$ into an inverted index, which is essentially a bipartite graph $G=(\mathcal{Q} \cup \mathcal{D}, E')$, where its vertices are either representing queries ($q \in \mathcal{Q}$) or data ($u \in \mathcal{D} $).
Then, they adapt the \minlogapa optimization problem to this context, by formulating the \bimlogapa problem, which seeks an ordering of the nodes in $\mathcal{D}$ to minimize 
\begin{equation}\label{eq:bmloga}
  \sum_{q \in \mathcal{Q}}  f_{\pi}(q,\mathcal{N}(q)) = \sum_{q \in \mathcal{Q}} \sum_{i=1}^{d_q -1} \log{ (\pi(u_i) -\pi(u_{i+1}))}. 
\end{equation}
%
%Vertices $u_1,\ldots,u_{d_{q}}$ correspond to the neighbors of $q \in V_\bot$ and are ordered such that $\pi(u_1) < \ldots < \pi(u_{d_{q}})$ holds.
%
%Similar to the \minlogapa problem from Section~\ref{sec:back}, \bimlogapa is inspired by the technique of delta-encoding and the objective is minimized when the gaps in the adjacency lists for $V_\bot$'s vertices are small.
%
Note that this problem is slightly different from~(\ref{eq:minloga}) because of the bipartivity of the graph.
They propose a recursive bisection heuristic denoted \recbis to tackle this minimization problem. %\bimlogapa.
Their heuristic adapts the well-known Kernighan-Lin~\cite{keli:70} and Fidduchia-Mattheyses~\cite{Fima:82} heuristics for graph partitioning to the problem of vertex reordering.
In a few words, Dhulipala \textit{et al.}'s heuristic starts from an initial partition of the set $\mathcal{D}$ into two equal-sized sets $\mathcal{D}_{1}$ and $\mathcal{D}_{2}$ (so a random balanced assignment). 
Assuming $\mathcal{D}$'s vertices are to be ordered between $\pi_{\ell}$ and $\pi_{r}$ in the final ordering $\pi$, vertices of $\mathcal{D}_{1}$ are in positions $\pi_{\ell},\ldots, \pi_{\ell}+\vert\mathcal{D}_{1} \vert - 1$ and vertices of $\mathcal{D}_{2}$ are in positions $\pi_{\ell}+\vert\mathcal{D}_{1} \vert,\ldots,\pi_{r}$.
The heuristic consists in exchanging vertices between the two partitions as long as they diminish the cost of the objective function.
Note that \recbis does not directly optimize equation~\eqref{eq:bmloga}, but aims to optimize the expected cost instead, which is equivalent to considering that the neighbors of $q$ are arranged regularly at equal distance from each other.
%
%That is, for each $q \in Q$, it is assumed that all of $q$'s neighbors in $D_1$ (respectively $D_2$) are located at equal distance from each other.
%
This is a simplification for the fact that the exact location of these vertices inside $\mathcal{D}_{1}$ and $\mathcal{D}_{2}$ are not yet known at that step.
Once the partition of $\mathcal{D}$ into $\mathcal{D}_{1}$ and $\mathcal{D}_{2}$ has been determined, the process is repeated recursively on the two inverted indexes $G_1=(\mathcal{Q} \cup \mathcal{D}_{1},E_1)$ and $G_2=(\mathcal{Q} \cup \mathcal{D}_{2},E_2)$.
The internal ordering of the vertices of $\mathcal{D}_{1}$ and $\mathcal{D}_{2}$ is therefore fixed in the subsequent recursive calls.

In~\cite{dkkops:16}, the authors also mention using recursive bisection to compress actual bipartite graphs, but there is no detail about the specifics of the process on such graphs.
%
%More generally, there has been a wealth of work on the question of inverted index compression, for a review see~\cite{pibiri2020techniques}.
%
The \bimlogapa problem focuses mostly on ordering properly vertices of the data group $\mathcal{D}$, which is relevant when compressing inverted indexes.
Our approach consists in ordering both the top ($V_\top$) and bottom ($V_\bot$) sets of nodes, so that if we represent the bipartite graph with adjacency lists of nodes in $  V_\top $, ordering $ V_\top $ nodes improves the compression by benefiting from delta encoding,
while ordering nodes in $ V_\bot $ improves the compression by benefiting from referencing.

\section{Dual reordering scheme\label{sec:ordings}} 

%To the best of our knowledge, the problem of graph compression in bipartite graphs has not been studied thoroughly before. 
%
%We will argue later on that permuting the vertices of the two sets together is not necessary ideal. 
%
%First, we permute the vertices of $V_\top$ and $V_\bot$ together and then compress the permuted graph.

In this section, we describe our approach to vertex reordering in bipartite graphs.
As suggested in Section~\ref{sec:back}, the reordering process in the case of bipartite graphs aims at reorganizing nodes in a way that would improve the property of similarity: we want vertices with close indexes to share many common neighbors.
%
%\footnote{}
%
Note that for bipartite data, locality is not relevant as a vertex is connected to vertices of the other group only.

We may choose to represent the graph either by storing adjacency lists of $ \bot $ nodes or those of $ \top $ nodes. 
Note that our goal is to improve compression regardless of what the vertices represent, so one may try both $V_\bot$ and $V_\top$ as the query set, and then keep the option yielding the best compression rate.
%
%~(as in the smallest compressed file).
%
%For example, a customer-product graph may best be represented by keeping only the purchases of each customer, when we frequently need to access purchases per user.
%
%In the vocabulary of inverted indexes, it means that one of the two vertex sets $V_\top$ or $V_\bot$ assumes the role of indexing vertices while the other set has the role of data entries.
%
%This option is best suited when one seeks to store the data efficiently without any specific application in mind.
%
Without loss of generality, we assume from now on that $V_\bot$ is the set of queries and that $V_\top$ is the set of data entries. %which main motivation is to maximize the impact of referencing.
That is, for each node $u \in V_\bot$, we store the list of its neighbors $\mathcal{N}(u)$ in $V_\top$.
Under this representation, all edges in the graph are stored exactly once.
We set that $\pi$ is the permutation ordering $V_\top$ nodes and $ \phi $ the permutation ordering $V_\bot$ nodes.

First, we describe how we design the $ \pi $ ordering on $V_\top$ nodes, which is essentially based on \recbis with a few improvements.
Then, our efforts focus on finding $ \phi $, \textit{i.e.} the ordering of $V_\bot$, with the purpose of maximizing the impact of referencing.

%In the following, we will see how to efficiently reorder the vertices for each vertex set so as to maximize two distinct objectives.
%The objective function used for $V_\top$'s reordering is discussed in Subsection~\ref{ssec:mingaps} and entails the issue of minimizing gaps between successive vertices in the adjacency lists.
%On the other hand, the objective function for $V_\bot$ discussed in Subsection~\ref{sssec:refreord} is based on the idea of vertex referencing described in the previous section. 

%The recursive-bisection algorithm due to Dhulipala \textit{et al.} can then be applied straightforwardly in order to generate an ordering of the data entries

%One can also consider a local-search based approach which randomly changes the ordering of two sets as a costlier alternative to improving their basic approach.

\subsection{Top nodes ordering to improve delta encoding \label{sssec:heur}}

$\top $ nodes (\textit{i.e.,} data nodes) are reordered with the recursive bisection heuristic \recbis described in Section~\ref{sssec:recbis}, as it has been proved to be very efficient for minimizing gaps between consecutive vertices in the adjacency lists~\cite{dkkops:16}.
In the same paper, the authors briefly mentioned a few  ideas to improve \recbis.
Here, we propose some modifications in the direction that they suggest and incorporate two mechanisms to \recbis to refine $ V_\top $ ordering $\pi$.

\begin{figure}[H]
    \centering

\begin{tikzpicture}

\draw (0,0) -- (4,0) -- (4,1) -- (0,1) -- (0,0);
\draw[ultra thick] (0,0) -- (0,1) -- (1,1) -- (1,0) -- (0,0);
\draw[ultra thick] (2,0) -- (2,1) -- (3,1) -- (3,0) -- (2,0);
\node[text width=1cm, font=\Large] at (0,1.5) {$V_1$};   
\node[text width=1cm, font=\normalsize, scale=0.7] at (0.42,0.5) {$min_1(q)$};   
\node[text width=1cm, font=\normalsize, scale=0.7] at (2.4,0.5) {$max_1(q)$};   

\draw (5,0) -- (9,0) -- (9,1) -- (5,1) -- (5,0);
\draw[ultra thick] (6,0) -- (6,1) -- (7,1) -- (7,0) -- (6,0);
\draw (7,0) -- (7,1) -- (8,1) -- (8,0) -- (7,0);
\node[text width=1cm, font=\Large] at (9.5,1.5) {$V_2$}; 
\node[text width=1cm, font=\normalsize, scale=0.7] at (6.42,0.5) {$min_2(q)$};  

\draw[-latex, thick] (4.3,1.6)--(0.7,1.05);
\draw[-latex, thick] (4.3,1.6)--(2.7,1.05);
\node[text width=1cm, font=\normalsize] at (4.9,1.7) {$q$};
\draw[-latex, thick] (4.7,1.6)--(6.3,1.05);

\draw [thick, decorate,decoration={brace,amplitude=5pt,mirror,raise=1ex}]
  (2.5,0) -- (6.5,0) node[midway,yshift=-2em]{gap};

\end{tikzpicture}
    
    \caption{An example of the partition swapping heuristic for a $\bot$ vertex $q$. If $V_1$ precedes $V_2$, the gap associated with $q$ is the distance between $max_1(q)$ and $min_2(q)$ \textit{i.e.}, 3.}
    \label{fig:heur_example}
\end{figure}
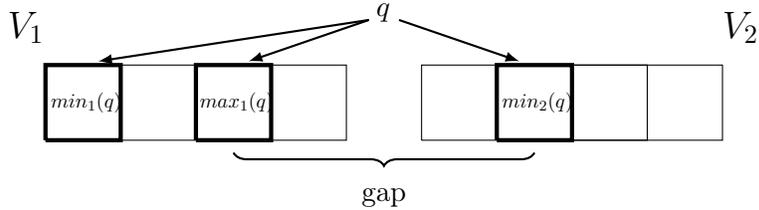

\paragraph{Partition swapping.}
%
%$ \top $ nodes are ordered using the recursive bisection described previously, as it is proven to be very efficient for this purpose.
%
%The authors of~\cite{dkkops:16} briefly hinted at a few ideas to improve it.
%
%Here, we take upon their hint and make changes on \recbis to refine $ V_\top $ ordering and improve compression.
%
The first mechanism is that of {\em partition swapping}.
Let us recall that in \recbis heuristic, a subset of nodes to reorder is partitioned into two sets $V_1$ and $V_2$ at each step.
This bisection aims at minimizing the inner gaps of $V_1$ and $V_2$.
In the original paper, the method always assumes that nodes of $V_1$ precede those of $ V_2 $ in the final ordering $ \pi $.
It is however possible that swapping $V_1$ and $ V_2 $ leads to a better ordering because it can potentially imply smaller gaps between the partitions.
We propose a linear-time heuristic to decide whether to apply the swap or not.
The heuristic relies on the cost function defined below.
\begin{definition}
For $q \in V_\bot$, $cost_{ij}(q)$ be the gap-cost of $q$ induced by the ordering $V_iV_j$ where $i,j \in \{1,2\}, i\neq j$.
\end{definition}
Note that the gap-cost is computed by summing over the $ \bot $ nodes (\textit{i.e.}, the queries),  as we aim to minimize the gaps between $ \top $ nodes, which are neighbors of a same $ \bot $ node.
At each step of the heuristic, we compute $\sum_{q \in V_\bot} cost_{12}(q)$ and $\sum_{q \in V_\bot} cost_{21}(q)$ and select the ordering between partitions that yields the lowest gap cost. 
To calculate the costs, we proceed as follows: given $V_1$ and $V_2$, we define $\zeta_i(d)$ to be the position of $d$ in the inner ordering of $V_i$ and thus a number between $1$ and $\vert V_i \vert$.
Once the inner orderings of  vertices inside $V_1$ and $V_2$ have been computed, we compute for each $q \in V_\bot$ and $i=1,2$ the quantities $max_{i}(q) = max \{ \zeta_i(d): (q,d) \in E, d \in V_i \}$ and $min_{i}(q) = min \{ \zeta_i(d): (q,d) \in E, d \in  V_i \}$.
These are respectively the largest and smallest $\zeta$ values of a neighbor of $q$ in $V_1$ or $V_2$, \textit{i.e.}, the first and last positions a neighbor of $q$ appears in $V_i$.
Observe now that any difference between $cost_{12}(q)$ and $cost_{21}(q)$ must be due to the gap between $q$'s last appearance in the preceding partition and its first appearance in the following partition, that is to say the gap between $max_1(q)$ and $min_2(q)$~for $V_1 V_2$ or $max_2(q)$ and $min_1(u)$~for $V_2 V_1$.
We illustrate this idea in Figure~\ref{fig:heur_example}.
The formula for $cost_{ij}(q)$ is formally given by
\begin{equation}
    cost_{ij}(q) =  \vert V_i \vert + min_j(q) - max_i(q). 
\end{equation}
%

%By the definition of the {\sc Gain} array, a vertex with high {\sc Gain} value is more preferable 

\paragraph{Partition flipping.}
To further improve the performance of the \recbis heuristic, we also incorporate an additional mechanism that we call {\em partition flipping}.
A flip in $V_i$ is the act of reversing the order of the vertices in a partition $V_i$ so that if the vertices are numbered from $1$ to $k$ then the vertex in position $k'$ swaps position with the vertex in $k-k'$ for all $ k' \in \llbracket 0,k-1 \rrbracket$.
Note that since flipping only reverses the order inside a partition, the gaps within a partition remain unchanged, but flipping allows to achieve smaller gaps between the partitions.
%
% It can be thought of as a method to correct some positionings achieved in previous steps.
%
For instance, considering the example of Figure~\ref{fig:heur_example}, if we flip partition $V_1$ then the gap related to node $q$ drops from 3 to 2, as the neighbor of $q$ previously ordered first in $V_1$ is now ordered last in $V_1$.\\

%Looking at the example of Figure~\ref{fig:heur_example} for example, if we flip partition $V_i$ then the gap for $q$ drops from 4 to 3.
%Overall, including flips, there is a total of 8 possible configurations to order $V^{1}_\top$ and $V^{2}_\top$.
%We again calculate the cost of each and pick the one incurring the smallest cost.

Adding both the swapping and flipping heuristics to \recbis implies a linear-time overhead, as we simply need to iterate over the current neighbors of each $q \in V_\bot$ to calculate the $min_{i}(q)$ and $max_{i}(q)$ values.
In practice, we observe that the overall time cost of these heuristics is negligible in comparison to the computation time of \recbis while these heuristics bring an improvement to the compression rate that is typically of the order of 1\%, as we will see in Section~\ref{subsec:recbis_heur}.

%\com{To be tested separately (and a few bits per vertex does not sound insignificant)} \ipcomment{I cannot test this but I think its not bad mentioning perhaps}

\subsection{Bottom nodes ordering to improve referencing \label{ssec:orii} \label{ssec:reftwo}}

We describe here our method to reorder $ \bot $ nodes  (\textit{i.e.,} query nodes according to our convention).
Let us recall that the referencing mechanism consists in allowing a vertex to encode part of its neighborhood implicitly by representing it as a bit array which contains 1 for neighbors in common with one of its predecessors, called the reference.
%
%Referencing can only occur if the two vertices are close enough to each other.
%
% with which $u$ shares many neighbors.
%
Consequently, the extent to which referencing is beneficial to compression largely depends on the ordering $\phi$ of the bottom nodes.
%
%As a consequence, the extent to which referencing impacts compression largely depends on the ordering of the indexing nodes.
% if vertex $u \in V_\bot$ is not preceded by a vertex with which it shares many neighbors, referencing will have no significant effect in reducing $u$'s storage costs.

%\paragraph{Reordering objectives.}
%
%The observation above leads us to reorder the indexing vertices ($V_\bot$) as well as vertices in adjacency lists ($V_\top$).
%
Efficient reordering approaches for $\top$ nodes focus primarily on orderings that minimize gaps between adjacent vertices stored in the adjacency lists through the \bimlogapa objective function, thus optimizing the delta encoding applied next.
%
%This is for instance the case of~\recbis .
%
Here, we are rather looking for referencing-friendly orderings for $\bot$ nodes, \textit{i.e.} orderings which decrease storage costs through the efficient use of referencing.
Note that orderings guaranteeing small gaps are likely to improve referencing too, however we argue that a $V_\bot$ reordering method should be explicitly designed to favor efficient referencing.
Ideally, we should express the reordering of $V_\bot$ with an objective function designed in the same way that \bimlogapa is to minimize gaps between vertices in $V_\top$.
Unfortunately, the cost of referencing depends not only on the referencing scheme considered, but also on the scheme used to encode integers, and therefore on the compression software used following the reordering phase, such as WebGraph or Zuckerli (see~\cite{bv04,vccg20}).
To explore the capabilities of the dual reordering scheme, we propose in what follows a software-agnostic optimization function for $ \bot $ nodes based on neighborhood similarity.

% lt: I am not really fan of the following paragraph...
%
%\ipadd{While we do not expect software-agnostic methods to exhibit the compression rates of their more specialized counterparts, we note that they can still act as indicative measures of how effective the dual residence scheme really is and showcase the usefulness of the scheme.}

%While an ordering guaranteeing small gaps is likely to improve referencing, we note that it seems counter-intuitive to rely an objective such as \bimlogapa for characterizing if an ordering is compression-friendly or not as they were not proposed for such a purpose.
%Indeed, thanks to our inverted index structure, we no longer care about minimizing gaps between consecutive vertices in the various adjacency lists.
%We henceforth believe that it makes more sense to reconsider the objective function being optimized and relate it more with how the referencing works.

%Ideally, in the second part, the objective used in the reordering problem should be directly related with the compression software used i.e.,  using Zuckerli would lead to a different ordering than using WebGraph.
%That is the case because the two softwares uses different encoding schemes for storing integers and thus yield different referencing costs.

%As can be sensed from the above, ignoring the bipartiteness of the graph to be compressed, and compressing both $V_\top$ and $V_\bot$ is something that will likely lessen the effect of referencing. That is so, because once again, we will have a single objective trying to satisfy two distinct optimization criterias.

\paragraph{\extshingle : Improving referencing through similarity. \label{sssec:shinglesplus}}

We define here the \extshingle heuristic, which takes inspiration from the shingle heuristic~\cite{chklmpr:09}.
Recall that the shingle is made to maximize the overlap between the neighborhoods of consecutive vertices.
So, it should be efficient to order $ \bot $ nodes in line with favoring referencing.
However, we observed in practice that it is not as effective as expected in this context.
One reason for this shortcoming is that many nodes can share the same shingle, thus creating buckets of similar values, and vertices in a same bucket are ordered arbitrarily.
As a result, it can happen that vertices with nearly identical neighborhoods are placed far apart in the same bucket, which can severely reduce the benefits of referencing.

The \extshingle heuristic orders the vertices within a bucket in a way that aims at maximizing the reference gain. 
The process works iteratively, by setting the position of one node in a bucket at each step.
More precisely, assuming that a bucket contains at step $i$ the ordered vertices $b_{1},\ldots, b_{i-1}$ while $  c_{i}, c_{i+1} \ldots, c_{k}$ vertices are still unordered,
the vertex from $c_{i}, \ldots, c_{k}$ selected to be in position $ b_{i} $ is the most similar to $b_{i-1}$.
Similarity is measured with the Jaccard index, we remind here that $ J(b,c) =  \vert \mathcal{N}(b) \cap \mathcal{N}(c) \vert / \vert \mathcal{N}(b) \cup \mathcal{N}(c) \vert$, which should ensure to have a good overlap between the neighborhoods of two consecutive $ \bot $ nodes and thus favor an effective referencing.
We describe the process more formally in Algorithm~\ref{algo:simref} for a bucket of size $k$ corresponding to vertices with a same shingle.

\begin{algorithm}[h]
  \begin{algorithmic}[1]
    \State {\bf Input:} randomly ordered table $c_{1},\ldots, c_{k}$ of vertices with the same shingle
    \State {\bf Output:} ordered table $b_{1},\ldots, b_{k}$ 
    \State {\bf Initialize} $ b_{1} = c_{1} $ 
    \For {$i = 2$ \TO $k$}
	    \State Compute $j_{max} = argmax_{j \in \llbracket  i,k \rrbracket } \left( J(b_{i-1},c_j) \right) $ 
	    \State Update $ b_i \leftarrow c_{j_{max}} $ , $ c_{j_{max}} \leftarrow c_i$ 
    \EndFor
  \end{algorithmic}
  \caption{Description of the \extshingle heuristic.}
  \label{algo:simref}
\end{algorithm}

As we will see in Section~\ref{subsec:dual_reorder}, the \extshingle heuristic brings improvements to the compression rate which vary depending on the dataset but can reach up to 6--7\%.
However, it is more computationally demanding than shingle is, as it adds a computation step which is quadratic in the size of the bucket.\footnote{A reader could notice that the general problem may be seen as an instance of the well-known Traveling Salesman Problem~(TSP) known to be NP-hard~\cite{jurr:95}:
given a weighted graph, the TSP aims at finding a route of minimum cost that starts and ends in the same vertex and passes through all other vertices exactly once.
Here the weight of an edge $(u,v)$ would correspond to $J(u,v)$, and \extshingle is a local search heuristic to approach a solution.}
Let us recall that the reordering time is not critical in general as this procedure is done offline and only once.

Nevertheless, it is possible to tune the trade-off between the computational cost of ordering nodes \textit{vs.} the quality of the ordering for referencing purposes.
%
%\com{ajout possible si place}
%
%Indeed, instead of using hash functions providing buckets, we can apply a similar heuristic on the entire $V_\bot$ set, as if only a single bucket existed, but the cost would be prohibitive.
%
%In that sense, partitioning $V_\bot$ into multiple buckets helps massively speed-up the entire process as the buckets should be of size significantly smaller than $ \vert V_\bot \vert $.
%
%If a bucket size exceeds a certain bound $D$, 
%
Indeed, the heuristic can be adapted to compute the similarity of a subset of the candidate set $\mathcal{C} = \{ c_{i}, \ldots, c_{j} \}$, and we briefly hint at two approaches for doing this. 
The first one is to sort the vertices in $\mathcal{C}$ by degree, and select nodes which have the closest degree to the one of $b_{i-1}$, as their Jaccard similarity is more likely to be high.
The second one is more elaborate, we generate a graph of nearest neighbors of the nodes in a bucket: we first compute for each vertex its $k$ most similar neighbors where $k$ is a parameter set by the user, then we create a graph $G_{s}$ where the edge $(u,v)$ exists iff $u$ is one of $v$'s most similar neighbors.
While considering $b_{i-1}$, its successor is selected among the available vertices in its neighborhood in $G_s$.
Computing the $k$-nearest neighbors can be done with approximate algorithms, which exhibit good performance in nearly linear time~\cite{docl:11}.

\section{Experiments \label{sec:exps}}

As described in Section~\ref{sec:ordings}, the dual reordering scheme that we propose consists in applying first the \recbis procedure with the swapping and flipping improvements to reorder $V_\top$ nodes by reducing the \bimlogapa objective.
Then we apply the \extshingle heuristic to reorder $V_\bot$ nodes to improve the referencing.
The code corresponding to this procedure is available here \url{https://bitbucket.org/IoannisPan/bipartitecompression/}\footnote{Note that the current implementation is supposed to be chained with Zuckerli compression method~\cite{vccg20}, itself available at \url{https://github.com/google/zuckerli}.}.
We describe in this section the experiments carried out to evaluate the efficiency of this scheme.

\subsection{Data and protocol}

We test our method on several massive bipartite graphs extracted from the KONECT network collection~\cite{konect}.\footnote{\url{http://konect.cc/}}
To evaluate the effect of the compression scheme in various contexts, we have selected several types of real-world datasets which can be represented by bipartite graphs.
We first consider a {\em social network}, containing actors starring in movies extracted from the Internet Movie Database.
Then we consider an {\em inverted index} which links together texts to the words that appear in them.
Finally, we select a group of networks which represent {\em human online activity}: editing activity on a Wikipedia or a Wiktionary, user tagging songs on Delicious, user listening to songs on LastFM.
%
%\ltcom{ [TODO] extend number and description of the datasets ; describe each family.} \ipcomment{i think it's fine like that, added}
%
When necessary, a dataset is pre-processed to eliminate multi-edges: for instance if a user has made several edits to a same Wikipedia page, it will be considered as one edge in the bipartite graph.
%\ltadd{
%
%Note also that in most datasets, a family of nodes is considered as a natural index while the other represent data entries.
%
%For example, when registering the edits made by a user of Wikipedia, the user is the index node while edits are entries.} 
%
The basic features of the datasets are described in Table~\ref{tab:bip_graphs}. 
%
%\ipcomment{I think we can delete these comments now}
%\ltcom{ I am sure about the number of nodes corresponding to users/pages is correct, but I am not so sure which family is considered as $V_\bot$ and which one is $V_\top$ with your heuristic, in the table I just put $V_\bot$ for the natural index and $V_\top$ fir the natural data entry. Can you please check?}
%\ipcomment{These look OK. Unfortunately, as I've also realized, the experiments should have been conducted with the transpose, and the actual results differ a bit. I had to modify slightly the code for this, and am looking at the effects of this change. }

%\ltadd{
\begin{table}[!h]
 \centering
  \caption{Number of nodes and edges of the bipartite graphs considered from~\cite{konect}.}
  \setlength{\tabcolsep}{4pt}
  \begin{tabular}{l|cc|cc|c}
    \toprule
     Graph  &  \multicolumn{2}{c|}{$V_\top$ nodes}   & \multicolumn{2}{c|}{$V_\bot$ nodes}   & edges \\ 
     \midrule
     {\tt imdb }			     & 303,617 & movies     &  896,302   & actors  & 	3,782,463 \\ 
     {\tt Reuters} 			     & 781,265 & texts    &  283,911   & words   & 60,569,726 \\				 
     {\tt lastfm-songs} 	     & 992     & users      & 1,084,620  & songs     & 4,413,834 \\
     {\tt Delicious~(user-tag)} & 833,081  & users      &  4,512,099 & tags & 81,989,133 \\
     {\tt En-wiktionary-edits } & 66,140  & users       &  5,826,113 & pages  & 27,120,425 \\
     {\tt Fr-wikipedia-edits}   & 	757,621 & users      & 8,870,762 & pages  & 52,950,008 \\
     {\tt De-wikipedia-edits}   &   1,025,084 & users   & 5,910,432 & pages  & 55,231,903 \\
    \bottomrule
  \end{tabular}
  \label{tab:bip_graphs}
\end{table}
%}

%Our methodology for compression is used as follows. 
%
%We treat each bipartite graph as an inverted index as discussed in Section~\ref{sec:ordings}. 
%
%$V_\top$'s vertices are always sorted with the recursive bisection algorithm of Dhulipala \textit{et al}~\cite{dkkops:16}, whereas those in $V_\bot$ are reordered with various reordering methods, depending on the experiment.

As a benchmark for comparison, we use the recursive bisection algorithm as proposed in~\cite{dkkops:16} on bipartite data.
Because \recbis is not described in details for bipartite graphs in this paper, we consider two options: i) in the first one, it orders $V_\top$ and $V_\bot$ as if they were one set of nodes \textit{i.e.}, the graph is processed in the same way as a unipartite graph would be, it is denoted \recbis-u; ii) in the second one, the recursive bisection method is applied to $V_\top$ and $V_\bot$ separately, in the same way as our dual scheme works, it is denoted \recbis-b.
As we will see, both options roughly yield the same compression rates (on average, \recbis-b outperforms \recbis-u by less than 1\%).
In all cases, the resulting orderings serve as inputs to a standard compression method.
The compression itself is achieved with Zuckerli~\cite{vccg20} with its default settings, as it is the current state-of-the-art solution  for graph compression.
As is usual in the domain, the compression quality is measured in average number of bits per edge in the compressed graph.
%
%The code is available online\footnote{\url{https://bitbucket.org/IoannisPan/bipartitecompression/}}, note that the code has been configured to work with Zuckerli compression software, which is also available online\footnote{\url{https://github.com/google/zuckerli}}.

\subsection{\recbis heuristics experiments}
\label{subsec:recbis_heur}

In the first set of experiments, we test 
%separately
the impact of the swapping and flipping heuristics from Section~\ref{sssec:heur} on the recursive bisection method.
The results with \recbis-b are summarized in Table~\ref{tab:heur_exps}.
We can observe that in all tested instances there is a gain on the average number of bits per edge required to store the graph, but these gains remain rather marginal as they are typically of the order of 1\% in the datasets that we have considered.
However, as the time cost of these heuristics is only a small fraction of the overall time required to run \recbis-b,
%
%, which implies that a user will certainly favor the memory saved over the time spent in ordering vertices.
%
we can recommend to systematically use the swapping and flipping heuristics when applying \recbis.

%Unfortunately, the overall increase is not large, however we note that for compression purposes every saved bit can be important.
%
%Furthermore, as the cost to employ the heuristic is only a mere fraction of the overall time required to run the algorithm in its entirety, we recommend that the proposed heuristic is always considered when applying \recbis.

\begin{table}[!h]
  \caption{Average number of bits per edge in the compressed representation using the standard \recbis-b algorithm derived from~\cite{dkkops:16} and the version of the algorithm with the swapping and flipping heuristics (S\&F).}
  \centering
  \setlength{\tabcolsep}{4pt}
  \begin{tabular}{l|ccc}
    \toprule
     Graph  &  \recbis-b  & \recbis-b+S\&F & \ \ \ gain (\%) \\ 
     \midrule
     {\tt imdb}  & 10.30 &  10.15 & 1.46\\ 
     {\tt Reuters} & 4.69 & 4.66 & 0.64\\ 			 
     {\tt lastfm-songs} &5.06 & 4.99 & 1.38\\
     {\tt Delicious~(user-tag)} & 6.86 & 6.82 & 0.58\\
     {\tt En-wiktionary-edits }  & 2.02 & 2.00 & 1.00\\
     {\tt Fr-wikipedia-edits} & 6.54 & 6.48 & 0.92\\	
     {\tt De-wikipedia-edits} &   8.48 & 8.39 & 1.06\\
    \bottomrule
  \end{tabular}
  \label{tab:heur_exps}
\end{table}

\subsection{Dual reordering scheme experiments}
\label{subsec:dual_reorder}

We now present experimental results to evaluate the practical efficiency of the whole dual reordering scheme.
The results are exhibited in Table~\ref{tab:zuck_graphs}.
The Natural ordering denotes the compression obtained using the initial ordering of the nodes in $V_\bot$ as produced by the data collection method from~\cite{konect} and \recbis on those of $V_\top$ and denotes a compression approach that does not make any attempt to optimize referencing.
\recbis-u corresponds to the baseline described in~\cite{dkkops:16}, where the vertices of $V_\bot$ and $V_\top$ are  reordered together as if the graph were unipartite.
The results with these two methods may be seen as standard approaches in the sense that they do not treat $V_\bot$ and $V_\top$ as different entities whereas the following two approaches do so. 
\recbis-b denotes the approach which applies \recbis separately on the sets of $V_\bot$ and $V_\top$.
Finally, our complete dual reordering scheme applying \recbis on $V_\top$ and \extshingle on $V_\bot$ is denoted \dual.
Note that we want to separate the improvements due to the S\&F heuristics from the improvement due to \extshingle, so in all uses of the \recbis algorithm, we apply the proposed heuristic from Section~\ref{sssec:heur}, which implies that the \recbis-b column is identical to the \recbis-b+S\&F heuristics column from Table~\ref{tab:heur_exps}.

%\ltcom{[Explain why we draw a line between left and right (I think it's SoA vs our methods) ; I think we did but check if we say which methods does the gain correspond to? DUAL vs. RecBis-bp/RecBis-up or natural?]}

\begin{table}[!h]
  \caption{Compression results using Zuckerli over different node orderings. The gain is the improvement from the \dual reordering scheme to \recbis-u (with the switching and flipping heuristics in all cases).}
  \setlength{\tabcolsep}{4pt}
  \centering
  \begin{tabular}{l|ccc|cc}
    \toprule
     Graph  &  Natural & \recbis-u~\cite{dkkops:16} &\recbis-b & \dual & gain (\%) \\ \midrule
     {\tt imdb} & 12.77 & 10.19 & 10.15 & 9.53 & 6.48 \\
     {\tt Reuters} & 4.71 & 4.67 & 4.66 & 4.66 & 0.21\\
     {\tt lastfm-songs} & 6.00 & 4.98 &  4.99 & 4.66 & 6.43 \\		
     {\tt Delicious~(user-tag)} & 7.53 & 6.82 &  6.82 &  6.69 & 1.91 \\ %\midrule
     {\tt En-wiktionary-edits} & 4.11 & 2.00 & 2.00 & 1.91 & 4.50\\
     {\tt Fr-wikipedia-edits} &  8.89 & 6.46 & 6.48  & 6.38 & 1.24\\				     	        
     {\tt De-wikipedia-edits} &  10.24 & 8.39 & 8.39 & 8.28 & 1.31 \\
    \bottomrule
  \end{tabular}
  \label{tab:zuck_graphs}
\end{table}

%The values in the table correspond to the average number of bits required per edge with each of these approaches using the Zuckerli compression software.

Unsurprisingly, the worst compression rates are obtained with the Natural orderings, as there is no specific effort to optimize referencing.
The performance of this method showcases the importance of reordering vertices to achieve efficient referencing.
The results obtained with \recbis-u and \recbis-b are nearly identical.
This observation stems from the fact that $ \top $ nodes (and respectively $ \bot $ nodes) are more similar to each other than they are to nodes of the other group, consequently \recbis-u tends to not intermix the two groups of nodes and thus orders nodes as \recbis-b does.
%
%Note that should the initial ordering for \recbis-ug be randomized such that vertices of $V_\bot$ and $V_\top$ intermix with each other, then \recbis-ug yields slightly worse results.
%
Most importantly, we observe that the compression rates obtained with the dual reordering scheme (\dual column) outperforms the other methods.
The gain depends on the dataset under consideration being lower than $2 \%$ for some datasets ({\tt Reuters}, {\tt Delicious}, {\tt Fr/De-wikipedia-edits}) but reaching up to $6-7 \%$ for others ({\tt imdb}, {\tt lastfm}), which is substantial in the domain of lossless graph compression.

A few additional remarks regarding these results guide us to look for further improvements.
First, the two ordering steps of the dual scheme can be iterated to compress further the graph until the orderings do not significantly change.
However, we observed that the improvements brought by iterating the process are marginal on the datasets under study (typically less than 1\%, not reported here).
Second, it is possible to switch the roles of $ \top $ and $ \bot $ nodes in the networks that we considered, as we are representing them as bipartite graphs without regard for what the nodes represent.
The results that we have shown here correspond to the choice yielding the best compression rates for each dataset. 
We can see in Table~\ref{tab:bip_graphs} that in all cases (except for {\tt Reuters}, which always yields poor gains), $|V_\top| < |V_\bot|$.
It seems to indicate that improving referencing yields better results when there is a larger choice of reference nodes to pick from.

From our perspective, the most important thing to draw from these experiments is that the \dual ordering scheme seems promising as it consistently outperforms other ordering methods, and that this mainly stems from the improvement made on the ordering $ \phi $ which targets referencing.
We can thus think of further developments to improve the referencing gains in bipartite graphs, as discussed in the next section.

%to take from these results is the fact that our strategy indeed works. The \extshingle heuristic which we designed to improve referencing consistently outperformed its competitors.

%As we can see, the results with \recbis-ug and \recbis-bp are nearly identical except for some slight differences in a few of the graphs.
%That can partly be explained by the fact, vertices in $V_\top$ are similar to those of $V_\bot$ and therefore it is expected that they would be grouped together by the swapping approach of \recbis.
%That is also helped by the fact, that in \recbis-ug the original ordering assumed during the initialization phase considers that the vertices of $V_\top$ precede those of $V_\bot$.
%This is something that already hints at the need to treat the the vertices of $V_\bot$ and $V_\top$ separately.

%Looking at the results in the \extshingle column, we see that ultimately it yields the best results.  The most noticeable drop is seen in graphs {\tt imdb} and {\tt lastfm(song)} where we notice drops of nearly $6-7 \%$.
%In the rest of the graphs, the drops are  usually in the range of $1-2 \%$, but an additional drop of around $1\%$ can be considered should we compare these results against the \recbis  method that does not use the proposed heuristic (compare against the Standard column in Table~\ref{tab:heur_exps}).

%%%%

\section{Discussion on further developments \label{sec:discus}}

\paragraph{Combining ordering heuristics.}
We have implemented the dual reordering scheme on bipartite graphs using a combination of \recbis for ordering $ \top $ nodes and \extshingle for ordering $ \bot $ nodes with encouraging results.
It is likely that other heuristic combinations can lead to better compression rates.
For instance, \extshingle is based on the Jaccard similarity, but other vertex similarity measures can be implemented, such as the Adamic-Adar index~\cite{adamic2003friends} or the Resource Allocation index~\cite{zhou2009predicting}.

The efficiency of the combinations is certainly data-dependent.
Understanding precisely why a dataset benefits more or less from a compression scheme originates from the particularities of the graph.
This point deserves deeper examination, and we think that the tools used in Zuckerli~\cite{vccg20} to investigate which parts of the compressed graph require the most bits is useful to pinpoint steps in the compression process that could be improved.

\paragraph{Referencing scheme.}
Another lead is to investigate the referencing scheme itself and possibly define a similarity metric based on this scheme.
We suggest here two ideas which can be interesting ways to improve the referencing scheme.

First, we can apply a \textit{post-processing reordering} on the $\phi$ ordering for $\bot$ nodes.
Supposing that we represent the fact that $u$ references $v$ by a directed arc from $u$ to $v$, then the references among $ \bot $ nodes form a forest: each vertex can only have one reference -- its parent -- and can be referenced by several nodes -- its children.
In addition, when vertex $u$ references another vertex $v$, the gap between $u$ and $v$  must be stored.
Given such a tree, we can create a new ordering $ \phi'$  where nodes of $ V_\bot $ in the same tree of the forest, are given consecutive ids. 
This reordering will prohibit vertices in different trees from alternating with each other in the final ordering and thus decrease the referencing gaps between them, thereby improving the final compression rate.

Another possibility is to allow {\em forward referencing}. 
With the usual referencing techniques discussed earlier, a node can only select its reference backwards, in the sense that we look for a reference in a window of nodes located before in the $\phi$ ordering.
By allowing a node to select its reference forward, we allow the referencing scheme to be more flexible, which can potentially improve the compression rates.
To illustrate this idea, let us consider a toy example of two nodes $u , v \in V_\bot$, $u$ has the following neighbors: $2, 4,\ldots, 10$, while $v$ has $1, 2, 3, \ldots 10$, in addition $u$ precedes $v$ in $\phi$.
With a standard referencing, we can only have $u \rightarrow v$ which leads to $1,3,\ldots, 9$ being residual neighbors  that are encoded separately with delta encoding.
As the neighbors of $v$ can be encoded more efficiently because they are consecutive, this reference would not be picked.
With the forward scheme, the reference $v \rightarrow u$ is also considered, and it would be efficient as $u$ could have its neighbors encoded with a simple bit array.
Note however, that this modification comes at a cost:  references no longer form a tree but a directed graph, and we must ensure that the directed graph of references is a directed acyclic graph, \textit{i.e.}, there is no cycle of references.

Preliminary results -- based on real data but only approximating the expected referencing gain -- indicate that these modifications could improve the compression rate up to a few percents.
To go further in this direction, it is necessary to select a specific compression method and to dive into its code to adjust the methods to the specifics of its implementation, which we leave to future investigations.

%We close the section by mentioning a small adjustment on the reordering of vertex set $V_\bot$ that can be applied once references between vertices have been chosen. 
%
%Our adjustment works by attempting to ignore those vertices between $u$ and $v$ that are not involved with either $u$ or $v$.
%
%More formally, let us consider the directed graph where edges $u \rightarrow v$ represent that $u$ chose to reference $v$.
%
%Notice that such a graph must necessarily be composed of many tree subgraphs.
%Assuming that ordering $\pi$ is used on the $V_\bot$'s vertices, we create an ordering $\pi'$ by assigning consecutive ids to vertices belonging to the same reference tree.
%
%Inside each tree the vertices are stored in accordance with their original ordering, i.e., if $\pi(u) < \pi(u')$ then $\pi'(u)< \pi'(u')$ holds so that the references are maintained as is.
%
%In practice, the maximum allowed gap for referencing is usually not very large, hence such a technique can only save a few bits per vertex.

%%%%

\section{Conclusion \label{sec:conc}}

In this work, we have examined the problem of lossless compression of bipartite graphs and proposed a dual reordering scheme of the vertices.
The central idea is to reorder the vertices of each partition with a different perspective in mind: either to optimize delta encoding or to maximize the effect of referencing, two techniques which are essential for standard compression methods.
We have shown empirically that this approach outperforms the classic single ordering methods, however the range of the improvement varies significantly depending on the dataset under study.
These encouraging results guided us to propose several leads for further improvements with the idea that reordering can be suited to specific datasets and specific referencing schemes.

%In addition, we have proposed a simple linear-time heuristic to improve the state-of-the-art ordering algorithm~\cite{dkkops:16} for minimizing gaps.

%For the future, we are going to continue in this direction and examine different objective functions in order to further utilize the mechanism of referencing.In particular, we would like to experiment with different notions of similarity, to see whether a metric different than Jaccard could lead to better performance.To that end, defining similarity based on the cost of referencing between two vertices sounds like the natural way to progress.

\subsection*{Acknowledgements}

We thank Fabrice Lécuyer and Matthieu Latapy for their proofreading of the manuscript.
This work was supported by the French National Agency of Research (ANR) by the Limass project (under grant ANR-19-CE23-0010).

\subsection*{Authors contribution statement}
M.~Danisch proposed the original idea of the paper, I.~Panagiotas realized the codes and experiments, I.~Panagiotas and L.~Tabourier contributed equally to the analysis and the writing of the paper.

\bibliographystyle{comnet}
\bibliography{references}

\begin{thebibliography}{00}

\bibitem{adamic2003friends}
Adamic, L.~A. {\&} Adar, E. (2003)  Friends and neighbors on the web. {\em
  Social networks}, \textbf{25}(3), 211--230.

\bibitem{aodr:09}
Apostolico, A. {\&} Drovandi, G. (2009)  Graph compression by BFS. {\em
  Algorithms}, \textbf{2}(3), 1031--1044.

\bibitem{beho:18}
Besta, M. {\&} Hoefler, T. (2018)  Survey and taxonomy of lossless graph
  compression and space-efficient graph representations. {\em arXiv preprint
  arXiv:1806.01799}.

\bibitem{b:15}
Block, P. (2015)  Reciprocity, transitivity, and the mysterious three-cycle.
  {\em Social Networks}, \textbf{40}, 163--173.

\bibitem{borsv:11}
Boldi, P., Rosa, M., Santini, M. {\&} Vigna, S. (2011)  Layered label
  propagation: A multiresolution coordinate-free ordering for compressing
  social networks. In {\em Proceedings of the 20th international conference on
  World Wide Web}, pages 587--596.

\bibitem{bv04}
Boldi, P. {\&} Vigna, S. (2004)  The webgraph framework I: compression
  techniques. In {\em Proceedings of the 13th international conference on World
  Wide Web}, pages 595--602.

\bibitem{bilana:09}
Brisaboa, N.~R., Ladra, S. {\&} Navarro, G. (2009)  k 2-trees for compact web
  graph representation. In {\em International symposium on string processing
  and information retrieval}, pages 18--30. Springer.

\bibitem{broder2000min}
Broder, A.~Z., Charikar, M., Frieze, A.~M. {\&} Mitzenmacher, M. (2000)
  Min-wise independent permutations. {\em Journal of Computer and System
  Sciences}, \textbf{60}(3), 630--659.

\bibitem{buce:08}
Buehrer, G. {\&} Chellapilla, K. (2008)  A scalable pattern mining approach to
  web graph compression with communities. In {\em Proceedings of the 2008
  international conference on web search and data mining}, pages 95--106.

\bibitem{chklmpr:09}
Chierichetti, F., Kumar, R., Lattanzi, S., Mitzenmacher, M., Panconesi, A. {\&}
  Raghavan, P. (2009)  On compressing social networks. In {\em Proceedings of
  the 15th ACM SIGKDD international conference on Knowledge discovery and data
  mining}, pages 219--228.

\bibitem{cana:10}
Claude, F. {\&} Navarro, G. (2010)  Fast and compact web graph representations.
  {\em ACM Transactions on the Web (TWEB)}, \textbf{4}(4), 1--31.

\bibitem{derara:06}
Delpratt, O., Rahman, N. {\&} Raman, R. (2006)  Engineering the LOUDS succinct
  tree representation. In {\em International Workshop on Experimental and
  Efficient Algorithms}, pages 134--145. Springer.

\bibitem{dkkops:16}
Dhulipala, L., Kabiljo, I., Karrer, B., Ottaviano, G., Pupyrev, S. {\&}
  Shalita, A. (2016)  Compressing graphs and indexes with recursive graph
  bisection. In {\em Proceedings of the 22nd ACM SIGKDD International
  Conference on Knowledge Discovery and Data Mining}, pages 1535--1544.

\bibitem{docl:11}
Dong, W., Charikar, M. {\&} Li, K. (2011)  Efficient k-nearest neighbor graph
  construction for generic similarity measures. In {\em Proceedings of the 20th
  international conference on World wide web}, pages 577--586.

\bibitem{Elias:75}
Elias, P. (1975)  Universal codeword sets and representations of the integers.
  {\em IEEE transactions on information theory}, \textbf{21}(2), 194--203.

\bibitem{Fima:82}
Fiduccia, C.~M. {\&} Mattheyses, R.~M. (1982)  A linear-time heuristic for
  improving network partitions. In {\em 19th design automation conference},
  pages 175--181. IEEE.

\bibitem{gabi:14}
Grabowski, S. {\&} Bieniecki, W. (2014)  Tight and simple web graph compression
  for forward and reverse neighbor queries. {\em Discrete Applied Mathematics},
  \textbf{163}, 298--306.

\bibitem{jurr:95}
J{\"u}nger, M., Reinelt, G. {\&} Rinaldi, G. (1995)  The traveling salesman
  problem. {\em Handbooks in operations research and management science},
  \textbf{7}, 225--330.

\bibitem{keli:70}
Kernighan, B.~W. {\&} Lin, S. (1970)  An efficient heuristic procedure for
  partitioning graphs. {\em The Bell system technical journal}, \textbf{49}(2),
  291--307.

\bibitem{konect}
Kunegis, J. (2013)  {KONECT} -- {The} {Koblenz} {Network} {Collection}. In {\em
  Proc. Int. Conf. on World Wide Web Companion}, pages 1343--1350.

\bibitem{lamo:00}
Larsson, N.~J.,  {\&} Moffat, A. (2000)  Off-line dictionary-based compression.
  {\em Proceedings of the IEEE}, \textbf{88}(11), 1722--1732.

\bibitem{lipasi:14}
Liakos, P., Papakonstantinopoulou, K. {\&} M.~Sioutis, M. (2014)  Pushing the
  envelope in graph compression. In {\em Proceedings of the 23rd ACM
  International Conference on Conference on Information and Knowledge
  Management}, pages 1549--1558.

\bibitem{moffat2019huffman}
Moffat, A. (2019)  Huffman coding. {\em ACM Computing Surveys (CSUR)},
  \textbf{52}(4), 1--35.

\bibitem{moffat1998arithmetic}
Moffat, A., Neal, R.~M. {\&} Witten, I.~H. (1998)  Arithmetic coding revisited.
  {\em ACM Transactions on Information Systems (TOIS)}, \textbf{16}(3),
  256--294.

\bibitem{pibiri2019fast}
Pibiri, G.~E., Petri, M. {\&} Moffat, A. (2019)  Fast dictionary-based
  compression for inverted indexes. In {\em Proceedings of the Twelfth ACM
  International Conference on Web Search and Data Mining}, pages 6--14.

\bibitem{pibiri2020techniques}
Pibiri, G.~E. {\&} Venturini, R.~R. (2020)  Techniques for inverted index
  compression. {\em ACM Computing Surveys (CSUR)}, \textbf{53}(6), 1--36.

\bibitem{rice1971adaptive}
Rice, R. {\&} Plaunt, J. (1971)  Adaptive variable-length coding for efficient
  compression of spacecraft television data. {\em IEEE Transactions on
  Communication Technology}, \textbf{19}(6), 889--897.

\bibitem{rozh:18}
Rossi, R.~A. {\&} Zhou, R. (2018)  Graphzip: a clique-based sparse graph
  compression method. {\em Journal of Big Data}, \textbf{5}(1), 1--14.

\bibitem{vccg20}
Versari, L., Comsa, I.-M., Conte, A. {\&} Grossi, R. (2020)  Zuckerli: A New
  Compressed Representation for Graphs. {\em IEEE Access}, \textbf{8},
  219233--219243.

\bibitem{zhou2009predicting}
Zhou, T., L{\"u}, L. {\&} Zhang, Y.-C. (2009)  Predicting missing links via
  local information. {\em The European Physical Journal B}, \textbf{71}(4),
  623--630.

\end{thebibliography}

\end{document}